\documentclass[12pt,psamsfonts,reqno]{amsart}

\theoremstyle{definition}

\theoremstyle{remark}

\newtheorem*{remark*}{Remark}

\newcommand{\vev}[1]{\left\langle #1 \right\rangle}
\newcommand{\ket}[1]{\left |  #1 \right \rangle}
\newcommand{\bra}[1]{\left \langle  #1 \right |}

\newcommand {\CalE} {\mathcal E}

\newcommand {\CalN} {\mathcal N}

\newcommand {\CalM} {\mathcal M}

\newcommand {\BR}   {\mathbb R}
\newcommand {\BZ}   {\mathbb Z}
\newcommand {\BC}   {\mathbb C}

\newcommand{\msS}{\mathscr{S}}

\newcommand{\sa}{\mathsf{a}}

\newcommand{\sw}{\mathsf{w}}
\newcommand{\sv}{\mathsf{v}}

\newcommand{\sA}{\mathsf{A}}

\newcommand{\sP}{\mathsf{P}}
\newcommand{\sQ}{\mathsf{Q}}

\newcommand{\sT}{\mathsf{T}}
\newcommand{\sV}{\mathsf{V}}
\newcommand{\sW}{\mathsf{W}}

\newcommand{\sY}{\mathsf{Y}}

\newcommand{\fq}{\mathfrak{q}}

\DeclareMathOperator{\Tr} {Tr}

\DeclareMathOperator{\rk} {rk}
\DeclareMathOperator{\vol}{vol}

\newcommand {\U} {\mathrm{U}}
\newcommand {\SU} {\mathrm{SU}}

\newcommand {\GL} {\mathrm{GL}}

\numberwithin{equation}{section}

\usepackage[margin=1in]{geometry}

\usepackage{amsmath}
\usepackage{amssymb}
\usepackage{amsxtra}
\usepackage{amscd}
\usepackage[mathscr]{euscript}

\usepackage[pagebackref=false]{hyperref}
\hypersetup{colorlinks = false, ,extension = notused,linkcolor = blue, anchorcolor = red,citecolor = blue,filecolor = red,pagecolor = red,urlcolor = blue, breaklinks=true}
\usepackage[numbers,sort&compress]{natbib}
\usepackage{hypernat}

\usepackage{iftex}
\ifxetex
        \usepackage{fontspec}
\else
\fi

\usepackage{float}
\usepackage{tikz-cd}
\usetikzlibrary{decorations.pathreplacing}
\usepackage[mathscr]{euscript}

\begin{document}

\title[Integrating over quiver variety and BPS/CFT correspondence]{Integrating over quiver variety\\[.5em] and\\[.5em] BPS/CFT correspondence}

\author{Taro Kimura}

%\address{Institut de Math\'ematiques de Bourgogne, Universit\'e de Bourgogne, France}
\address{Institut de Math\'ematiques de Bourgogne, UMR 5584, CNRS, Universit\'e Bourgogne Franche-Comt\'e, F-21000 Dijon, France
}

 \begin{abstract}
  We show the vertex operator formalism for the quiver gauge theory partition function and the $qq$-character of highest-weight module on quiver, both associated with the integral over the quiver variety.
 \end{abstract}

\maketitle 

\tableofcontents

\parskip=4pt

\section{Introduction}\label{sec:intro}

Let $\CalM_{k,n}$ be the instanton moduli space of $k$-instanton configuration in 4d $\SU(n)$ gauge theory on $\BR^4 \cong \BC^2$.
This moduli space plays a significant role in the study of 4d gauge theory.
In particular, we are interested in the volume of the moduli space, which gives rise to the contribution to the partition function of 4d $\CalN = 2$ supersymmetric gauge theory.
The total partition function is given as the summation over the instanton sectors~\cite{Nekrasov:2002qd},
\begin{align}
 Z = \sum_{k=0}^\infty \fq^k \, Z_k
\end{align}
where the instanton fugacity $\fq$ is related to the the complexified coupling constant $\displaystyle \tau = \frac{\theta}{2\pi} + \iota \frac{4\pi}{g^2}$ by $\fq = \exp \left( 2 \pi \iota \tau \right)$ with the imaginary unit $\iota = \sqrt{-1}$.
$\theta$ and $g^2$ are the theta angle and the gauge coupling constant, appearing in the Lagrangian.
In fact, the $k$-instanton contribution is given as the volume of the instanton moduli space
\begin{align}
 Z_k = \vol(\CalM_{k,n})
 \, .
\end{align}
Since the instanton moduli space is non-compact, we need a regularization scheme to evaluate the volume:
The equivariant integral provides a proper regularized volume of the instanton moduli space, which ends up with the contour integral, called the Losev--Moore--Nekrasov--Shatashvili (LMNS) formula~\cite{Moore:1997dj,Losev:1997tp,Lossev:1997bz},%
\footnote{%
This formula is extended to BCD-type~\cite{Nekrasov:2004vw,Marino:2004cn} and supergroup gauge theory~\cite{Kimura:2019msw}.
}
\begin{align}
 \vol(\CalM_{k,n}) = \frac{1}{k!} \frac{\epsilon^k}{\epsilon_1^k \epsilon_2^k} \oint \prod_{a = 1}^k \frac{d \phi_a}{2 \pi \iota} \prod_{\alpha = 1}^n \frac{1}{(\phi_a - \sa_\alpha)(\phi_a - \sa_\alpha + \epsilon)}
 \prod_{a < b}^k \frac{\phi_{ab}^2 (\phi_{ab}^2 - \epsilon^2)}{(\phi_{ab}^2 - \epsilon_1^2)(\phi_{ab}^2 - \epsilon_2^2)}
 \, .
 \label{eq:LMNS}
\end{align}
where $\phi_{ab} = \phi_a - \phi_b$.
$(\phi_a)_{a = 1, \ldots, k}$ and $(\sa_\alpha)_{\alpha = 1, \ldots, n}$ are the sets of the equivariant parameters for $\GL(k,\BC)$ and GL($n, \BC$) acting on the instanton moduli space.
The equivariant parameters for $\mathrm{Spin}(4)$ acting as the Lorentz transformation for $\BR^4 \cong \BC^2$ is given by $(\epsilon_1, \epsilon_2)$ and $\epsilon = \epsilon_1 + \epsilon_2$.
The partition function depends on these equivariant parameters, and the Seiberg--Witten prepotential is consequently reproduced by asymptotic behavior of the partition function in the limit $\epsilon_{1,2} \to 0$.

The instanton moduli space for $\BC^2$ is an example of the quiver variety corresponding to $\widehat{A}_0$ quiver~\cite{Nakajima1994,Nakajima:1998DM}, and from this point of view, it is natural to consider the equivariant volume of generic quiver varieties.
In this paper, we focus on the quiver gauge theory on $\BC^2$, and consider similar integrals to the LMNS formula~\eqref{eq:LMNS} associated to the quiver, which provide the corresponding gauge theory partition functions.
We point out that the multi-variable integral is concisely expressed as a correlation function of vertex operators with the free field realization depending on the quiver structure.
We remark that this vertex operator is nothing but the operator introduced in Refs.~\cite{Kimura:2015rgi,Kimura:2016dys,Kimura:2017hez} to construct the quiver W-algebra.
Such a connection between the gauge theory observable and the vertex operators is now called the BPS/CFT correspondence~\cite{Nekrasov:2015wsu,Nekrasov:2016qym,Nekrasov:2016ydq,Nekrasov:2017rqy,Nekrasov:2017gzb}.
An earlier remarkable example is the equivalence between the $\SU(2)$ instanton partition function and the Liouville conformal block, a.k.a., the AGT relation~\cite{Alday:2009aq}.
The recent progress on the supersymmetric localization~\cite{Pestun:2007rz,Pestun:2016zxk} allows us to obtain a lot of exact results on the gauge theory side, and many examples of the correspondence can be now checked so far based on these results.

Another gauge theory observable associated to the quiver variety is the $qq$-character~\cite{Nekrasov:2015wsu}, which is a two parameter deformation of the character of representations constructed on the quiver.
Realization of the representations using the quiver variety is now established~\cite{Nakajima:1999}, and with this realization, the $qq$-character is constructed by integrating over the quiver variety.
We show that the vertex operator formalism discussed in this paper is also applicable to this $qq$-character integral construction.
It has been known that the $qq$-character plays a role of the generating current of the quiver W-algebra~\cite{Kimura:2015rgi,Kimura:2016dys,Kimura:2017hez}, which implies the operator formalism provides a still formal, but concise closed formula for arbitrary generating current of the W-algebra.

The remaining part of the paper is organized as follows:
In Sec.~\ref{sec:VO}, we introduce the vertex operators associated with the quiver.
We explicitly show the operator product expansion (OPE) factors between them based on their free field realization.
In Sec.~\ref{sec:Z}, we consider the partition function of 5d $\CalN = 1$ quiver gauge theory defined on $\BC^2 \times S^1$, which is the K-theoretic analog of the equivariant volume of the instanton moduli space.
We show that the gauge theory partition function is expressed as a correlator of the vertex operators associated to the quiver.
In Sec.~\ref{sec:qq-ch}, we consider the $qq$-character based on the integral formula.
We show that the integral measure is reproduced by the OPE factors between the vertex operators, and the highest-weight module is generated by the Weyl orbit generating operator.
In Sec.~\ref{sec:6d}, we consider 6d $\CalN = (1,0)$ gauge theory on $\BC^2 \times T^2$, which yields an elliptic analog of these results.
We show that there are two approaches to the elliptic theory.
One is to consider the torus correlation function instead of the correlation function on a cylinder $\BC^\times$.
The other is deformation of the vertex operators providing the elliptic OPE factors.

%Sec.~\ref{sec:discussion}

\subsubsection*{Acknowledgements}

We would like to thank Saebyeok Jeong, Alexei Morozov, Nikita Nekrasov, Vasily Pestun, Maxim Zabzine, Yegor Zenkevich for useful discussions and comments.
A part of the results in this paper was presented in \href{https://indico.ipmu.jp/indico/event/168/}{Representation theory, gauge theory, and integrable systems} at Kavli IPMU, February 2019, and \href{https://conferences.cirm-math.fr/2058.html}{BPS/CFT correspondence} at CIRM, September 2019.
We are grateful to the organizers for giving the opportunity to present our work in a stimulating atmosphere.
This work was supported by the French ``Investissements d'Avenir'' program, project ISITE-BFC (No.~ANR-15-IDEX-0003), JSPS Grant-in-Aid for Scientific Research (No.~JP17K18090), the MEXT-Supported Program for the Strategic Research Foundation at Private Universities ``Topological Science'' (No.~S1511006), JSPS Grant-in-Aid for Scientific Research on Innovative Areas ``Topological Materials Science'' (No.~JP15H05855), and ``Discrete Geometric Analysis for Materials Design'' (No.~JP17H06462).

\section{Vertex operators}\label{sec:VO}

\subsection{Quiver}

We follow the convention used in~\cite{Kimura:2015rgi}.
We denote a quiver,%
\footnote{%
We consider simply-laced quivers here.
We can generalize the result in this paper to the fractional (non-simply-laced; symmetrizable) quiver using the formalism introduced in~\cite{Kimura:2017hez}.
}
consisting of nodes and edges, by $\Gamma = (\Gamma_0,\Gamma_1)$ where $\Gamma_0$ is a set of nodes, and $\Gamma_1$ is a set of edges.
The rank of quiver is given as $\rk \Gamma = |\Gamma_0|$.
For such a quiver, we define the mass deformed $q$-Cartan matrix
\begin{align}
 c_{ij} = (1 + q^{-1}) \delta_{ij} - \sum_{e:i \to j} \mu_e^{-1} - \sum_{e:j \to i} \mu_e q^{-1}
 \, ,
\end{align}
where $q \in \BC^\times$, and $(\mu_e)_{e \in \Gamma_1}$ is a set of $\BC^\times$-valued parameters assigned to each edge.
In the gauge theory setup, the former one is associated with the equivariant parameter for Spin($4$) action $q = q_1 q_2$ where
\begin{align}
 (q_1, q_2) := (e^{\epsilon_1}, e^{\epsilon_2}) \in \U(1)^2 \subset \mathrm{Spin}(4)
 \, .
\end{align}
The latter ones are identified with the multiplicative bifundamental mass parameters, which are given from the additive bifundamental mass $(m_e)_{e \in \Gamma_1}$ as $\mu_e = e^{m_e} \in \BC^\times$.

We denote the Adams operation of the $q$-Cartan matrix, replaced with the degree $n$ monomials, by
\begin{align}
 c_{ij}^{[n]} = (1 + q^{-n}) \delta_{ij} - \sum_{e:i \to j} \mu_e^{-n} - \sum_{e:j \to i} \mu_e^n q^{-n}
 \, .
\end{align}
We remark that it is reduced to the ordinary quiver Cartan matrix in {\em the classical limit} $n \to 0$,
\begin{align}
 c_{ij}^{[0]} = 2 \delta_{ij} - \#\{e:i \to j\} - \#\{e:j \to i\}
 \, .
\end{align}

\subsection{Vertex operators}\label{sec:VOA}

We introduce the vertex operators associated to the quiver $\Gamma$ using the free field oscillators.
We then summarize the OPE factors between them, which will be used in the following argument.

\subsubsection{$\sA$-operator}

We define the $\sA$-operator for each node of the quiver $i \in \Gamma_0$,
\begin{align}
 \sA_i(x) = \ : \exp \left( a_{i,0} - \kappa_i \log x + \sum_{n \neq 0} a_{i,n} \, x^{-n} \right) : 
\end{align}
where $: \ :$ is the normal ordering symbol, and the free field oscillators obey the commutation relation
\begin{align}
 \Big[ a_{i,n},\, a_{j,m} \Big] = - \frac{1}{n} (1 - q_1^n)(1 - q_2^n) \, c_{ji}^{[n]} \, \delta_{n+m,0}
 \, .
 \label{eq:aa_rel}
\end{align}
This oscillator has an explicit realization using formal parameters $(t_{i,n})_{i \in \Gamma_0, n = 1,\ldots,\infty}$,
\begin{align}
 (n > 0) \quad
 a_{i,-n} = (1 - q_1^n)(1 - q_2^n) t_{i,n}
 \, , \quad
 a_{i,n} = - \frac{1}{n} c_{ji}^{[n]} \frac{\partial}{\partial t_{j,n}}
 \, , \quad
 a_{i,0} = - \log \fq_i
 \, ,
\end{align}
where $(\fq_i)_{i \in \Gamma_0}$ is the gauge coupling constant, and the formal parameter $(t_{i,n})_{i \in \Gamma_0, n = 1,\ldots,\infty}$ is identified with the coupling constant for the generic Casimir operator~\cite{Nekrasov:2003rj,Nakajima:2003uh,Marshakov:2006ii}.
The zero mode coefficient $(\kappa_i)_{i \in \Gamma_0} \in \mathbb{Z}^{|\Gamma_0|}$ is to be identified with the Chern--Simons level of the corresponding gauge theory.
This vertex operator is slightly modified from the original definition by Frenkel--Reshetikhin~\cite{Frenkel:1997} and also from~\cite{Kimura:2015rgi}, used in the construction of the $q$-deformed W-algebras.

\subsubsection{$\sY$-operator}

We define the $\sY$-operator
\begin{align}
 \sY_i(x) = \ : \exp \left( y_{i,0} - \tilde{c}_{ji}^{[0]} \kappa_j \log x + \sum_{n \neq 0} y_{i,n} \, x^{-n} \right) :
 \, ,
\end{align}
where the free field oscillator obeys the commutation relation
\begin{align}
 \Big[ y_{i,n}, \, y_{j,m} \Big] = - \frac{1}{n} (1 - q_1^n)(1 - q_2^n) \, \tilde{c}_{ij}^{[-n]} \, \delta_{n+m,0}
\end{align}
It has the following realization using the formal parameters $(t_{i,n})_{i \in \Gamma_0, n = 1,\ldots,\infty}$,
\begin{align}
 (n > 0) \quad
 y_{i,-n} = (1 - q_1^n)(1 - q_2^n) \tilde{c}_{ij}^{[-n]} t_{j,n}
 \, , \quad
 y_{i,n} = - \frac{1}{n} \frac{\partial}{\partial t_{i,n}}
 \, , \quad
 y_{i,0} = - \tilde{c}_{ji}^{[0]} \log \fq_j
\end{align}
where we denote the inverse of the $q$-Cartan matrix by $\tilde{c}_{ij}^{[n]}$.%
\footnote{%
For the affine quiver, the quiver Cartan matrix is not invertible in particular for $n = 0$ since $\det (c_{ij}^{[0]}) = 0$.
In this case, we should deal with the zero mode separately.
See~\cite{Kimura:2015rgi} for details.
}

We remark that the $a$- and $y$-oscillators behave as the root and the weight since they are converted to each other through the $q$-Cartan matrix
\begin{align}
 a_{i,n} = \sum_{j \in \Gamma_0} y_{j,n} \, c_{ji}^{[n]}
 \, ,
\end{align}
which implies the relation between the $\sA$- and $\sY$-operators as follows:
\begin{align}
 \sA_i(x) = \ : \frac{\displaystyle \sY_i(x) \sY_i(qx)}{\displaystyle \prod_{e:i \to j} \sY_j(\mu_e q^{-1} x) \prod_{e:j \to i} \sY_j(\mu_e^{-1} x) }:
 \, .
 \label{eq:AY_rel}
\end{align}
Such an interpretation is relevant to the construction of representations on the quiver~\cite{Nakajima:1999}, which will be discussed in Sec.~\ref{sec:qq-ch}.

\subsubsection{$\sV$-operator}

We define the $\sV$-operator
\begin{align}
 \sV_i(x) = \ : \exp \left( \sum_{n \neq 0} v_{i,n} \, x^{-n} \right) :
 \, ,
\end{align}
which has the realization
\begin{align}
 (n > 0) \quad
 v_{i,-n} = - \tilde{c}_{ji}^{[n]} t_{j,n}
 \, , \quad
 v_{i,n} = \frac{1}{n} \frac{1}{(1 - q_1^n)(1 - q_2^n)} \frac{\partial}{\partial t_{i,n}}
 \, .
 \label{eq:v_osc}
\end{align}
We remark the relation to the $y$-oscillator
\begin{align}
 v_{i,n} = - \frac{1}{(1 - q_1^n)(1 - q_2^n)} \, y_{i,n}
 \, ,
\end{align}
and thus this $\sV$-operator is related to the weight associated with the quiver.

\subsubsection{OPE factors}\label{sec:OPEs}

In order to write down the OPE factors, we define a rational function
\begin{align}
 \msS(x) =
 \frac{(1 - q_1 x)(1 - q_2 x)}{(1 - x)(1 - q x)} =
 \exp \left( \sum_{n=1}^\infty \frac{1}{n} (1 - q_1^n)(1 - q_2^n) x^n \right)
 \, ,
\end{align}
which obeys the reflection relation
\begin{align}
 \msS(x^{-1}) = \msS(q^{-1} x)
 \label{eq:S_ref}
 \, .
\end{align}
Then the OPE factors between the vertex operators are given as follows:\\

\paragraph{$\sA\sA$ OPE}
From the commutation relation \eqref{eq:aa_rel}, we obtain
\begin{align}
 \sA_i(x) \sA_j(x') = \ :\sA_i(x) \sA_j(x'): \times
 \begin{cases}
  \displaystyle
  \msS \left( \frac{x'}{x} \right)^{-1} \msS \left( \frac{x}{x'} \right)^{-1}
  & (i = j)
  \\[1em] \displaystyle
  \msS \left( \mu_{e} q^{-1} \frac{x'}{x} \right)
  = \msS \left( \mu_{e}^{-1} \frac{x}{x'} \right)
  & (\text{s.t.} \ e: i \to j)
  \\[1em] \displaystyle
  \msS \left( \mu_{e}^{-1} \frac{x'}{x} \right)
  = \msS \left( \mu_{e} q^{-1} \frac{x}{x'} \right)
  & (\text{s.t.} \ e: j \to i)
  \\[1em] \displaystyle
  1 & (\text{otherwise})
 \end{cases}
 \label{eq:AA_rel}
\end{align}

\paragraph{$\sY\sA$ OPE}
Since the $y$- and $a$-oscillators obey the commutation relation
\begin{align}
 \Big[ y_{i,n} , \, a_{j,m} \Big] = - \frac{1}{n} (1 - q_1^n) (1 - q_2^n) \, \delta_{ij} \, \delta_{n+m,0}
 \, ,
\end{align}
we obtain
\begin{align}
 \sY_i(x) \sA_j(x') = \sA_j(x') \sY_i(x) = \msS\left( \frac{x'}{x} \right)^{-\delta_{ij}} :\sY_i(x) \sA_j(x'):
 \, ,
 \label{eq:YA_rel}
\end{align}       
for $x'/x \neq 1, q^{-1}$.

\paragraph{$\sV\sA$ OPE}
The $v$- and $a$-oscillators obey the commutation relation
\begin{align}
 \Big[ v_{i,n} , \, a_{j,m} \Big] = \frac{1}{n} \, \delta_{ij} \, \delta_{n+m,0}
 \, ,
\end{align}
thus we obtain
\begin{subequations}\label{eq:VA_rel}
 \begin{align}
  \sV_i(x) \sA_j(x') & = \left( 1 - \frac{x'}{x} \right)^{-\delta_{ij}} :\sV_i(x) \sA_j(x'): \, ,
  \\
  \sA_j(x') \sV_i(x) & = \left( 1 - \frac{x}{x'} \right)^{-\delta_{ij}} :\sV_i(x) \sA_j(x'): \, .
 \end{align}
\end{subequations}

\section{Gauge theory partition function}\label{sec:Z}

We consider the partition function of 5d $\CalN = 1$ quiver gauge theory on $\BC^2 \times S^1$, and show that it is expressed as a correlation function of the vertex operators introduced in Sec.~\ref{sec:VO}.

\subsection{$A_1$ quiver}

\subsubsection{$\CalN = 1$ pure Yang--Mills theory}

The partition function of SU($n$) Yang--Mills theory ($A_1$ quiver) is given as the sum over the instanton contributions:
\begin{align}
 Z = \sum_{k=0}^\infty \fq^k \, Z_k
 \, .
\end{align}
The $k$-instanton contribution leads to the following contour integral,
\begin{align}
 Z_k = \frac{1}{k!} \left( \frac{1 - q}{(1 - q_1)(1 - q_2)} \right)^k
 \oint \prod_{a=1}^k \frac{d \phi_a}{2 \pi \iota \phi_a}
% \phi_a^\kappa
 \frac{\phi_a^\kappa}{\sP(\phi_a) \widetilde{\sP}(q \phi_a)}
 \prod_{a \neq b}^k \msS\left( \frac{\phi_a}{\phi_b} \right)^{-1}
% \prod_{a = 1}^k \prod_{\alpha = 1}^n \left( 1 - \frac{\nu_\alpha}{\phi_a} \right)^{-1} \left( 1 - q \frac{\phi_a}{\nu_\alpha} \right)^{-1}
 \, ,
 \label{eq:qLMNS}
\end{align}
which is the $q$-analog of the LMNS formula \eqref{eq:LMNS} for 4d $\CalN = 2$ gauge theory.
Here $\sP(x)$ and $\widetilde{\sP}(x)$ are the gauge polynomials given by
\begin{align}
 \sP(x) = \prod_{\alpha=1}^n \left( 1 - \frac{\nu_\alpha}{x} \right)
 \, , \qquad
 \widetilde{\sP}(x) = \prod_{\alpha=1}^n \left( 1 - \frac{x}{\nu_\alpha} \right)
 \, ,
\end{align}
where $(\nu_\alpha)_{\alpha = 1,\ldots, n}$ is a set of the multiplicative Coulomb moduli parameters, given by $\nu_\alpha = e^{\sa_\alpha} \in \BC^\times$, and $\kappa \in \BZ$ is the Chern--Simons level.
The $k$-instanton contribution for 5d theory is not directly interpreted as the equivariant volume of the instanton moduli space, but it is given as the twisted Witten index of the supersymmetric quantum mechanics on $S^1$, whose Hilbert space is given by the instanton moduli space $\CalM_{k,n}$.
We remark that the integral \eqref{eq:qLMNS} is a multi-variable contour integral, so that we have to fix the integral contour properly.
A modern characterization of such a contour integral is by the Jeffrey--Kirwan residue prescription.
See, for example, \cite{Benini:2013xpa} for the explantion in a similar context.
The resultant residue is consistent with the fixed point in the instanton moduli space under the equivariant action in this case.

For the latter simplicity we shift the coupling constant
\begin{align}
 \fq \ \longrightarrow \ \frac{1 - q}{(1 - q_1)(1 - q_2)} \, \fq
 \, ,
 \label{eq:coupling_shift}
\end{align}
then, using the OPE factors shown in Sec.~\ref{sec:OPEs}, we obtain the correlator formula for the partition function,
\begin{align}
 \fq^k \, Z_k
 = \frac{1}{k!} \oint \prod_{a=1}^k \frac{d \phi_a}{2 \pi \iota \phi_a}
 \bra{\sV^{(n)}} \prod_{a=1}^k \sA(\phi_a)^{-1} \ket{\sV^{(n)}}
 / \vev{\sV^{(n)} \mid \sV^{(n)}}
\end{align}
where the $\sV$ state $\ket{\sV^{(n)}}$ and its dual $\bra{\sV^{(n)}}$ are defined with the $\sV$ operator,
\begin{align}
 \ket{\sV^{(n)}} = \ :\prod_{\alpha=1}^n \sV(\nu_\alpha): \ket{0}
 \, , \quad
 \bra{\sV^{(n)}} = \bra{0} :\prod_{\alpha=1}^n \sV(q^{-1} \nu_\alpha):
 \, .
\end{align}
The vacuum state $\ket{0}$ and its dual $\bra{0}$ are annihilated by the negative and positive modes,
\begin{align}
 (n>0) \qquad
 \frac{\partial}{\partial t_n} \ket{0} = \bra{0} t_n = 0
 \, .
\end{align}
The normalization factor $\vev{\sV^{(n)} \mid \sV^{(n)}}$ cancels the OPE factors between the $\sV$-operators%
\footnote{%
We remark that this normalization factor is slightly similar, but different from the perturbative contribution to the gauge theory partition function, written using the $q$-double $\Gamma$-function,
\begin{align}
 \Gamma_2(x;q_1,q_2) = \prod_{n,n'=0}^\infty \left( 1 - x q_1^n q_2^{n'} \right)^{-1} = \exp \left( \sum_{m=1}^\infty \frac{1}{m} \frac{1}{(1 - q_1^m)(1 - q_2^m)} x^m \right)
 \, .
\end{align}
}
\begin{align}
 \vev{\sV^{(n)} \mid \sV^{(n)}} = \exp \left( \sum_{m=1}^\infty \frac{1}{m} \frac{1}{(1 - q_1^m)(1 - q_2^m)} \frac{1}{1 + q^{-m}} \sum_{\alpha,\beta}^n \frac{\nu_\beta}{\nu_\alpha} \right)
 \, .
\end{align}

Furthermore, we define the $\sW$-operator, which is a charge associated to the root operator,
\begin{align}
 \sW = \oint \frac{d \phi}{2 \pi \iota \phi} \, \sA(\phi)^{-1}
 \, .
\end{align}
Then the normalized partition function has a formal, but concise expression
\begin{align}
 \tilde{Z} & = \sum_{k=0}^\infty \frac{1}{k!} \bra{\sV^{(n)}} \sW^k \ket{\sV^{(n)}} %/ \vev{\sV^{(n)} \mid \sV^{(n)}}
 = \bra{\sV^{(n)}} e^{\sW} \ket{\sV^{(n)}} %/ \vev{\sV^{(n)} \mid \sV^{(n)}}
 \, .
 \label{eq:tildeZ}
\end{align}
We have several remarks on this formula.
First, a proper regularization would be necessary on the product $\sW^k$, which may provides a singularity due to the operator collision.
A possible way of regularization is to deform the vertex operator with a regularization parameter, and then take this parameter to zero after the computation.%
\footnote{%
We thank Nikita Nekrasov for pointing out this issue.
}

Next is the relation of the operator $\sW$ to the screening charge in 2d CFT, which is defined as an integral of the screening current.
The screening charge is a similar formal operator, which does not make sense by itself, but plays an important role in derivation of an integral formula for the CFT correlation function, a.k.a., the Dotsenko--Fateev integral formula~\cite{Dotsenko:1984nm,Dotsenko:1984ad}.
In fact, the $\sA$-operator used here is directly related to the screening current for the $q$-deformation of W-algebras by $q$-difference operation~\cite{Frenkel:1997,Kimura:2015rgi,Kimura:2016dys,Kimura:2017hez}.
This suggests a possible interpretation of the $\sW$-operator as an alternative screening charge in the $q$-deformed setup.

Another remark is the similarity of the expression \eqref{eq:tildeZ} to the several formulas in the literatures, e.g., the Fourier transform of the gauge theory partition function, known as the dual partition function~\cite{Nekrasov:2003rj}, the $\widehat{\textrm{W}}$-operator representation of the matrix integral~\cite{Morozov:2009xk}, and the partition function as the norm of the Gaiotto--Whittaker state~\cite{Gaiotto:2009ma}.
Actually such an expression is often found in the context of the integrable hierarchy as the corresponding $\tau$-function.
It would be interesting to pursue the connection between the formula presented here and other similar formulas.

\subsubsection{$\CalN = 1$ SQCD}

We then consider 5d $\CalN = 1$ SQCD, which has additional $n^\text{f}$ fundamental and $n^\text{af}$ antifundamental hypermultiplets.
In this case, the $k$-instanton contribution has additional factors as follows:
\begin{align}
 Z_k = \frac{1}{k!} \left( \frac{1 - q}{(1 - q_1)(1 - q_2)} \right)^k
 \oint \prod_{a=1}^k \frac{d \phi_a}{2 \pi \iota \phi_a} \,
 \phi_a^\kappa \,
 \frac{\sQ(\phi_a) \widetilde{\sQ}(q\phi_a)}{\sP(\phi_a) \widetilde{\sP}(q \phi_a)}
 \prod_{a \neq b}^k \msS\left( \frac{\phi_a}{\phi_b} \right)^{-1}
\end{align}
where we define the matter polynomials
\begin{align}
 \sQ(x) = \prod_{f=1}^{n^\text{f}} \left( 1 - \frac{\mu_f}{x} \right)
 \, , \qquad
 \widetilde{\sQ}(x) = \prod_{f=1}^{n^\text{af}} \left( 1 - \frac{x}{\tilde{\mu}_f} \right)
\end{align}
with the sets of the multiplicative fundamental and antifundamental mass parameters, $(\mu_f)_{f = 1,\ldots, n^\text{f}}$ and $(\tilde{\mu}_f)_{f = 1, \ldots, n^\text{af}}$.

In a similar way to the previous case, under the coupling shift \eqref{eq:coupling_shift}, we obtain 
\begin{align}
 \fq^k \, Z_k
 = \frac{1}{k!} \oint \prod_{a=1}^k \frac{d \phi_a}{2 \pi \iota \phi_a}
 \langle~ \sV^{(n,n^\text{af})} \mid \prod_{a=1}^k \sA(\phi_a)^{-1} \mid \sV^{(n,n^\text{f})} ~\rangle
 / \langle~ \sV^{(n,n^\text{af})} \mid \sV^{(n,n^\text{f})} ~\rangle
 \label{eq:Zk_correlator}
\end{align}
with the modified $\sV$-state,
\begin{subequations}
\begin{align}
 \mid \sV^{(n,n^\text{f})} ~\rangle
 & = \ :\prod_{\alpha=1}^n \sV(\nu_\alpha) \prod_{f=1}^{n^\text{f}} \sV(\mu_f)^{-1}: \ket{0}
 \, , \\
 \langle~ \sV^{(n,n^\text{af})} \mid
 & = \bra{0} :\prod_{\alpha=1}^n \sV(q^{-1} \nu_\alpha) \prod_{f=1}^{n^\text{af}} \sV(q^{-1}\tilde{\mu}_f)^{-1}:
 \, .
\end{align}
\end{subequations}
The additional contribution of the (anti)fundamental matter is imposed by this modification of the $\sV$-state.
Then, summing up all the instanton sectors, we obtain the normalized partition function
\begin{align}
 \tilde{Z} & = \sum_{k=0}^\infty \frac{1}{k!} \langle~\sV^{(n,n^\text{af})} \mid \sW^k \mid \sV^{(n,n^\text{f})} ~\rangle
 = \langle~\sV^{(n,n^\text{af})} \mid e^{\sW} \mid \sV^{(n,n^\text{f})} ~\rangle
 \, .
 \label{eq:Z_correlator}
\end{align}
We remark that one can use the same $\sW$-operator as far as considering $A_1$ quiver gauge theory, and the matter content dependence appears only in the $\sV$-state.%
\footnote{%
The theory with the adjoint matter is classified into $\widehat{A}_0$ quiver theory.
See Sec.~\ref{sec:hatA0} for details.
}

\subsection{Quiver gauge theory}\label{sec:generic_quiver}

We study 5d $\CalN = 1$ $\Gamma$-quiver gauge theory on $\BC^2 \times S^1$.
We define the dimension vectors,
$\underline{k} = (k_i)_{i \in \Gamma_0}$, $\underline{n} = (n_i)_{i \in \Gamma_0}$, $\underline{n}^\text{(a)f} = (n_i^\text{(a)f})_{i \in \Gamma_0}$, which characterize the instanton moduli space denoted by $\CalM_{\underline{k},\underline{n}}$.
The gauge theory partition function is given as the summation over topological sectors characterized by the dimension vector,
\begin{align}
 Z = \sum_{\underline{k}} \fq^{\underline{k}} \, Z_{\underline{k}}
\end{align}
where we use the convention
\begin{align}
 \sum_{\underline{k}} = \sum_{k_1=0}^{\infty} \sum_{k_2=0}^\infty \cdots \sum_{k_{\rk \Gamma}=0}^\infty
 \, , \qquad
 \fq^{\underline{k}} = \prod_{i \in \Gamma_0} \fq_i^{k_i}
 \, .
\end{align}
The $\underline{k}$-instanton contribution to the partition function has the following contour integral form: 
\begin{align}
 Z_{\underline{k}} & = \prod_{i \in \Gamma_0} \frac{1}{k_i!}
 \oint \prod_{i \in \Gamma_0}
 \prod_{a=1}^{k_i}
 \frac{d \phi_{i,a}}{2 \pi \iota \phi_{i,a}} \,
 z_i(\phi_i; \nu_i,\mu_i, \tilde{\mu}_i)
 \prod_{e:i \to j} \prod_{a=1}^{k_i} \prod_{b=1}^{k_j} \msS \left( \mu_e q^{-1} \frac{\phi_{j,b}}{\phi_{i,a}} \right)
 \prod_{e:j \to i} \prod_{a=1}^{k_i} \prod_{b=1}^{k_j} \msS \left( \mu_e^{-1} \frac{\phi_{i,a}}{\phi_{j,b}} \right)
 \label{eq:Zk_quiver}
\end{align}
with the building block for each node $i \in \Gamma_0$
\begin{align}
 z_i(\phi_i; \nu_i,\mu_i, \tilde{\mu}_i) =
 \prod_{a=1}^{k_i}
 \left[
 \phi_{i,a}^{\kappa_i} \frac{ \sQ_i(\phi_{i,a}) \widetilde{\sQ}_i(q\phi_{i,a})}{\sP_i(\phi_{i,a}) \widetilde{\sP}_i(q \phi_{i,a})}
 \prod_{e:i \to j} \mathsf{P}_j(\mu_e^{-1} \phi_{i,a})
 \prod_{e:j \to i} \widetilde{\mathsf{P}}_j(\mu_e^{-1} q \phi_{i,a}) 
 \right]
 \prod_{a \neq b}^{k_i} \msS\left( \frac{\phi_{i,a}}{\phi_{i,b}} \right)^{-1}
 \, ,
\end{align}
and the shift of the coupling constant
\begin{align}
 \fq_i \ \longrightarrow \ \frac{1 - q}{(1 - q_1)(1 - q_2)} \, \fq_i
 \, .
 \label{eq:coupling_shift_quiv}
\end{align}
We define the gauge polynomials and the matter polynomials associated to each gauge node,
\begin{subequations}
\begin{align}
 \sP_i(x) & = \prod_{\alpha=1}^{n_i} \left( 1 - \frac{\nu_{i,\alpha}}{x} \right)
 \, , \qquad
 \widetilde{\sP}_i(x) = \prod_{\alpha=1}^{n_i} \left( 1 - \frac{x}{\nu_{i,\alpha}} \right)
 \, , \\
 \sQ_i(x) & = \prod_{f=1}^{n_i^\text{f}} \left( 1 - \frac{\mu_{i,f}}{x} \right)
 \, , \qquad
 \widetilde{\sQ}_i(x) = \prod_{f=1}^{n_i^\text{af}} \left( 1 - \frac{x}{\tilde{\mu}_{i,f}} \right)
 \, ,
\end{align}
\end{subequations}
with the multiplicative Coulomb moduli and the (anti)fundamental mass parameters,
\begin{align}
 \nu_i = (\nu_{i,\alpha})_{\alpha = 1,\ldots,n_i}
 \, , \qquad
 \mu_i = (\mu_{i,f})_{f = 1,\ldots,n_i^\text{f}}
 \, , \qquad
 \tilde{\mu}_i = (\tilde{\mu}_{i,f})_{f = 1,\ldots,n_i^\text{af}}
 \, .
\end{align}
In this case, we obtain the following correlator representation using the OPE factors shown in Sec.~\ref{sec:OPEs}:
\begin{align}
 Z_{\underline{k}} = \prod_{i \in \Gamma_0} \frac{1}{k_i!}
 \oint \prod_{i \in \Gamma_0}
 \prod_{a=1}^{k_i} \frac{d \phi_{i,a}}{2 \pi \iota \phi_{i,a}} 
 \langle~ \sV^{(\underline{n},\underline{n}^\text{af})} \mid \prod_{i \in \Gamma_0} \prod_{a=1}^{k_i} \sA_i(\phi_{i,a})^{-1} \mid \sV^{(\underline{n},\underline{n}^\text{f})} ~\rangle
 / \langle~ \sV^{(\underline{n},\underline{n}^\text{af})} \mid \sV^{(\underline{n},\underline{n}^\text{f})} ~\rangle
\end{align}
with the $\sV$-state
\begin{subequations}
\begin{align}
 \mid \sV^{(\underline{n},\underline{n}^\text{f})} ~\rangle
 & = \ :
 \prod_{i \in \Gamma_0}
 \left[
 \prod_{\alpha=1}^{n_i}
 \left(
 \sV_i(\nu_{i,\alpha})
 \prod_{e:i \to j}
 \sV_j(\mu_e \nu_{i,\alpha})^{-1} 
 \right)
 \prod_{f=1}^{n_i^\text{f}} \sV_i(\mu_{i,f})^{-1}
 \right]
 : \ket{0}
 \, , \\
 \langle~\sV^{(\underline{n},\underline{n}^\text{af})} \mid
 & = \bra{0} :
 \prod_{i \in \Gamma_0}
 \left[
 \prod_{\alpha=1}^{n_i}
 \left(
 \sV_i(\nu_{i,\alpha} q^{-1})
 \prod_{e:j \to i}
 \sV_j(\mu_e q^{-1}\nu_{i,\alpha})^{-1}
 \right)
 \prod_{f=1}^{n_i^\text{af}} \sV_i(\tilde{\mu}_{i,f} q^{-1})^{-1}
 \right]
 :
 \, .
\end{align}
\end{subequations}
Namely, all the factors appearing in the contour integral \eqref{eq:Zk_quiver} are reproduced by the OPE factors between the vertex operators associated to the quiver $\Gamma$.
Then, introducing the $\sW$-operator for each node $i \in \Gamma_0$
\begin{align}
 \sW_i = \oint \frac{d \phi}{2 \pi \iota \phi} \, \sA_i(\phi)^{-1}
 \, ,
 \label{eq:W_op}
\end{align}
the summation over the instanton sectors is given as
\begin{align}
 \tilde{Z} = \sum_{\underline{k}} 
 \langle~ \sV^{(\underline{n},\underline{n}^\text{af})} \mid \prod_{i \in \Gamma_0} \frac{\sW_i^{k_i}}{k_i!} \mid \sV^{(\underline{n},\underline{n}^\text{f})} ~\rangle
 =  \langle~ \sV^{(\underline{n},\underline{n}^\text{af})} \mid \prod_{i \in \Gamma_0} e^{\sW_i} \mid \sV^{(\underline{n},\underline{n}^\text{f})} ~\rangle
 \, .
\end{align}
This formalism is available for generic quiver, not restricted to the finite-type Dynkin quiver, but also the affine and hyperbolic quivers.

We remark that the vertex operators used in this paper are also used in another formalism~\cite{Kimura:2015rgi,Kimura:2016dys,Kimura:2017hez}, where the screening charge, given as the the screening current integral, plays a central role in the construction of the partition function instead of the $\sA$-operator.
The explicit relation between these two formulations is not yet obvious, whereas a clue would be that the $q$-difference of the screening current gives rise to the $\sA$-operator, as mentioned before.
We would discuss this issue in a future research.

\subsubsection{$\widehat{A}_0$ quiver}\label{sec:hatA0}

Let us consider $\widehat{A}_0$ quiver (5d $\CalN = 1^*$ $\SU(n)$ gauge theory), which is the simplest example of the affine quivers.
In this case, the $q$-Cartan matrix is given by
\begin{align}
 c = 1 + q^{-1} - \mu^{-1} - \mu q^{-1} = (1 - \mu^{-1})(1 - \mu q^{-1})
 \, ,
\end{align}
with the multiplicative adjoint mass $\mu = e^{m_\text{adj}} \in \BC^\times$.
In the classical limit, this $q$-Cartan matrix is reduced to $c^{[0]} = 0$, so that not invertible.
We now assume $c \neq 0 \iff \mu \neq 1, q$, and it becomes invertible after the $q$-deformation.

In this case, the $k$-instanton contribution is given by
\begin{align}
 Z_k = \frac{1}{k!} \oint \prod_{a=1}^k \frac{d \phi_a}{2 \pi \iota \phi_a} \frac{\sP(\mu^{-1} \phi_a) \widetilde{\sP}(\mu^{-1} q \phi_a)}{\sP(\phi_a) \widetilde{\sP}(q \phi_a)} \prod_{a \neq b}^k \msS \left( \frac{\phi_a}{\phi_b} \right)^{-1} \msS \left( \mu^{-1} \frac{\phi_a}{\phi_b} \right)
 \, ,
\end{align}
which is written using the vertex operators as follows:
\begin{align}
 \fq^k \, Z_k
 = \frac{1}{k!} \oint \prod_{a=1}^k \frac{d \phi_a}{2 \pi \iota \phi_a}
 \langle~ \sV^{(n,n)} \mid \prod_{a=1}^k \sA(\phi_a)^{-1} \mid \sV^{(n,n)} ~\rangle
 / \langle~ \sV^{(n,n)} \mid \sV^{(n,n)} ~\rangle
 \, .
\end{align}
The $\sV$-state is now defined
\begin{subequations}
\begin{align}
 \mid \sV^{(n,n)} ~\rangle
 & = \ :\prod_{\alpha=1}^n \sV(\nu_\alpha) \sV(\mu \nu_\alpha)^{-1}: \ket{0}
 \, , \\
 \langle~ \sV^{(n,n)} \mid
 & = \bra{0} :\prod_{\alpha=1}^n \sV( \nu_\alpha q^{-1}) \sV(\mu q^{-1} \nu_\alpha)^{-1}:
 \, ,
\end{align}
\end{subequations}
and the normalized partition function is given as the summation over the instanton sectors
\begin{align}
 \tilde{Z} & = \sum_{k=0}^\infty \frac{1}{k!} \langle~\sV^{(n,n)} \mid \sW^k \mid \sV^{(n,n)} ~\rangle
 = \langle~\sV^{(n,n)} \mid e^{\sW} \mid \sV^{(n,n)} ~\rangle
 \, ,
\end{align}
where the $\sW$-operator is the integral of the $\sA$-operator associated with $\widehat{A}_0$ quiver as in the previous case.

\section{$qq$-character integral formula}\label{sec:qq-ch}

The $qq$-character is the doubly quantum deformation of the character of representations constructed on the quiver $\Gamma$.
It was shown in~\cite{Nekrasov:2015wsu} that the $qq$-character has a formula based on the integration over the quiver variety associated to the quiver $\Gamma$.
We apply the operator formalism to the $qq$-character based on this integral formula.

We define the dimension vectors, $\underline{\sv} = (\sv_i)_{i \in \Gamma_0}$, $\underline{\sw} = (\sw_i)_{i \in \Gamma_0}$, characterizing the representation on the quiver~\cite{Nakajima:1999}.
Let $\underline{v} = (v_{i,a})_{i \in \Gamma_0,a=1,\ldots,\sv_i}$, $\underline{w} = (w_{i,\alpha})_{i \in \Gamma_0, \alpha = 1,\ldots,\sw_i}$ be the equivariant parameters for $\displaystyle \GL(\underline{\sv}) = \prod_{i \in \Gamma_0} \GL(\sv_i,\BC)$ and $\displaystyle \GL(\underline{\sw}) = \prod_{i \in \Gamma_0} \GL(\sw_i, \BC)$ action.
Then we introduce {\em the highest-weight operator},
\begin{align}
 \sY_{\underline{\sw}}(\underline{w})
 = \ :\prod_{i \in \Gamma_0} \prod_{\alpha=1}^{\sw_i} \sY_i(w_{i,\alpha}):
 \, .
 \label{eq:hw}
\end{align}
The $qq$-character is given as the summation over $\underline{\sv} \in \BZ_{>0}^{\rk \Gamma}$,
\begin{align}
 \sT_{\underline{\sw}}(\underline{w})
 & = \sum_{\underline{\sv}} % \fq^{\underline{\sv}} \,
 \sT_{\underline{\sw}, \underline{\sv}}(\underline{w})
 \, ,
 \label{eq:sum_v}
\end{align}
where each contribution $\sT_{\underline{\sw}, \underline{\sv}}(\underline{w})$ corresponds to the integral over the quiver variety $\CalM_{\underline{\sv}, \underline{\sw}}(\Gamma)$,
\begin{align}
 \sT_{\underline{\sw}, \underline{\sv}}(\underline{w})
 & = \prod_{i \in \Gamma_0} \frac{1}{\sv_i!}
 \oint \prod_{i \in \Gamma_0} \prod_{a = 1}^{\sv_i} \frac{d v_{i,a}}{2 \pi \iota v_{i,a}}
 \prod_{a \neq b}^{\sv_i} \msS \left( \frac{v_{i,b}}{v_{i,a}} \right)^{-1}
 \prod_{e:i \to j} \prod_{a=1}^{\sv_i} \prod_{b=1}^{\sv_j} \msS\left( \mu_e q^{-1} \frac{v_{j,b}}{v_{i,a}} \right)
 \prod_{e:j \to i} \prod_{a=1}^{\sv_i} \prod_{b=1}^{\sv_j} \msS\left( \mu_e^{-1} \frac{v_{i,a}}{v_{j,b}} \right)
 \nonumber \\
 & \hspace{10em} \times
 \prod_{a=1}^{\sv_i} \prod_{\alpha=1}^{\sw_i} \msS \left( \frac{v_{i,a}}{w_{i,\alpha}} \right)
 : \sY_{\underline{\sw}}(\underline{w})
 \prod_{a=1}^{\sv_i} \sA_i(v_{i,a})^{-1} 
 :
 \, .
\end{align}
We again shift the coupling constant \eqref{eq:coupling_shift_quiv}, which is absorbed by the normalization constant of the $\sA$-operator.

It turns out that all the $\msS$-factors appearing in the integral are precisely reproduced by the OPE factors of the $\sA$- and $\sY$-operators shown in Sec.~\ref{sec:OPEs}:
\begin{align}
 \sT_{\underline{\sw}, \underline{\sv}}(\underline{w})
 & = \oint \prod_{i \in \Gamma_0} \prod_{a = 1}^{\sv_i} \frac{d v_{i,a}}{2 \pi \iota v_{i,a}}
 \sY_{\underline{\sw}}(\underline{w})
 \prod_{i \in \Gamma_0} \prod_{a=1}^{\sv_i} \sA_i(v_{i,a})^{-1}
 \, .
 \label{eq:T_operator}
\end{align}
Then we obtain a formal, but concise formula for generic $qq$-character using the $\sW$-operator~\eqref{eq:W_op},
\begin{align}
 \sT_{\underline{\sw}}(\underline{w})
 & = \sum_{\underline{\sv}} %\fq^{\underline{\sv}} \,
 \sY_{\underline{\sw}}(\underline{w})
 \prod_{i \in \Gamma_0} \frac{\sW_i^{\sv_i}}{\sv_i!}
 = \sY_{\underline{\sw}}(\underline{w}) \prod_{i \in \Gamma_0} e^{\sW_i}
 \, .
 \label{eq:T_operator_full}
\end{align}
We remark that the $qq$-character is an operator acting on the Fock space generated by $(t_{i,n}, \partial_{t_{i,n}})_{i \in \Gamma_0, n = 1,\ldots, \infty}$, so that it is not expressed as a free field correlator.

From the representation theoretical point of view, the $qq$-character formula~\eqref{eq:T_operator_full} is interpreted as follows:
The operator defined in \eqref{eq:hw} is the highest-weight, and the sum over the dimension vector $\underline{\sv}$~\eqref{eq:sum_v} corresponds to the sum over the corresponding Weyl orbit.
The operator $\displaystyle \prod_{i \in \Gamma_0} e^{\sW_i}$ in the formula plays a role of the Weyl orbit generating operator.

We remark that the $qq$-character $\sT_{\underline{\sw}}(\underline{w})$ is now an operator acting on the Fock space, which becomes a pole-free regular function after taking the gauge theory average.
In the operator formalism, the regularity of the $qq$-character is rephrased as the commutativity with the screening charge associated with the quiver~\cite{Kimura:2015rgi,Kimura:2016dys,Kimura:2017hez}.
It would be worth studying the regularity and the commutativity in terms of the $\sW$-operator since it also has an essential connection with the screening charge in the quiver W-algebra formalism.
We leave this issue for a future study.

\section{6d $\CalN = (1,0)$ theory}\label{sec:6d}

In this Section, we show that the formalism shown above is naturally generalized to the 6d $\CalN=(1,0)$ theory compactified on a torus, $\BC^2 \times T^2$.
Let $\tau$ be the modulus of the torus on which the gauge theory is compactified.
We define the elliptic nome
\begin{align}
 p = e^{2 \pi \iota \tau} \in \BC^\times
 \, ,
\end{align}
and the theta function
\begin{align}
 \theta(x;p) = (x;p)_\infty (px^{-1};p)_{\infty}
 \, .
\end{align}
We remark the relation
\begin{align}
 \theta(x^{-1};p) = (-x^{-1}) \theta(x;p)
\end{align}
is essentially equivalent to $1 - x^{-1} = (-x^{-1})(1 - x)$.

\subsection{Partition function}

We consider $A_1$ quiver theory for simplicity.
We can similarly formulate generic quiver gauge theory based on the same argument in Sec.~\ref{sec:generic_quiver}.
See also~\cite{Kimura:2016dys} for related discussions.

The gauge theory partition function is given as the instanton sum
\begin{align}
 Z = \sum_{k=0}^\infty \fq^k \, Z_k
 \, .
\end{align}
The $k$-instanton contribution for 6d SU($n$) theory with $n^\text{f}$ and $n^\text{af}$ (anti)fundamental hypermultiplets is given by
\begin{align}
 Z_k = \frac{1}{k!} \left( \frac{\theta(q;p)}{\theta(q_1;p)\theta(q_2;p)} \right)^k \oint \prod_{a=1}^k \frac{d \phi_a}{2 \pi \iota \phi_a}
 \frac{\sQ(\phi_a) \widetilde{\sQ}(q \phi_a)}{\sP(\phi_a) \widetilde{\sP}(q \phi_a)}
 \prod_{a \neq b}^k \msS \left( \frac{\phi_a}{\phi_b} \right)^{-1}
\end{align}
where all the factors are replaced with the elliptic functions,
\begin{subequations}
\begin{align}
 \msS(x) = \msS(q^{-1} x^{-1}) = \frac{\theta(q_1 x;p)\theta(q_2 x;p)}{\theta(x;p) \theta(qx;p)} = \exp \left( \sum_{n \neq 0} \frac{1}{n} \frac{(1 - q_1^n)(1 - q_2^n)}{1 - p^n} x^n \right)
 \, ,
 \label{eq:elliptic_S}
\end{align}
 \begin{align}
  \sP(x) & = \prod_{\alpha = 1}^n \theta\left( \frac{\nu_\alpha}{x};p \right)
  \, , \qquad
  \widetilde{\sP}(x) = \prod_{\alpha = 1}^n \theta\left( \frac{x}{\nu_\alpha};p \right)
  \, , \\
  \sQ(x) & = \prod_{f = 1}^{n^\text{f}} \theta \left( \frac{\mu_f}{x};p \right)
  \, , \qquad
  \widetilde{\sQ}(x) = \prod_{f = 1}^{n^\text{af}} \theta \left( \frac{x}{\tilde{\mu}_f};p \right)
  \, .
 \end{align}
\end{subequations}
We remark that we should impose the anomaly free condition in order that the partition function possesses a proper modular property.

\subsubsection{Trace formula}

Let $L_0$ be the energy operator defined
\begin{align}
 L_0 = \sum_{m=1}^\infty m \, t_{m} \frac{\partial}{\partial t_m}
 \, .
\end{align}
Under the shift \eqref{eq:coupling_shift}, the $k$-instanton contribution is written as the trace over the Fock space generated by $(t_n, \partial_{t_n})_{n = 1,\ldots,\infty}$,
\begin{align}
 \fq^k \, Z_k
 = \frac{1}{k!} \oint \prod_{a=1}^k \frac{d \phi_a}{2 \pi \iota \phi_a}
 \Tr \left[ p^{L_0} \,
 \sV^{(n,n^\text{af})}
 \left( \prod_{a=1}^k \sA(\phi_a)^{-1} \right)
 \sV^{(n,n^\text{f})}
 \right] /
 \Tr \left[ p^{L_0} \,
 \sV^{(n,n^\text{af})} \sV^{(n,n^\text{f})}
 \right]
\end{align}
where we define
\begin{align}
 \sV^{(n,n^\text{f})} 
 & = \ :\prod_{\alpha=1}^n \sV(\nu_\alpha) \prod_{f=1}^{n^\text{f}} \sV(\mu_f)^{-1}: 
 \, , \quad
 \sV^{(n,n^\text{af})}
 = \ :\prod_{\alpha=1}^n \sV(q^{-1} \nu_\alpha) \prod_{f=1}^{n^\text{af}} \sV(q^{-1}\tilde{\mu}_f)^{-1}:
 \, .
\end{align}
Then, the normalized partition function is given by
\begin{align}
 \tilde{Z} & = \sum_{k=0}^\infty \frac{1}{k!} \Tr \left[ p^{L_0} \, \sV^{(n,n^\text{af})} \, \sW^k \, \sV^{(n,n^\text{f})} \right]
 = \Tr \left[ p^{L_0} \, \sV^{(n,n^\text{af})} \, e^{\sW} \, \sV^{(n,n^\text{f})} \right]
 \, .
\end{align}
This 6d formula is interpreted as a correlator on a torus $T^2$, or elliptic curve $\CalE_p = \BC^\times/p^\BZ = \BC / (\BZ \oplus \tau \BZ)$, while the 5d formula~\eqref{eq:Z_correlator} is a correlator on a cylinder $\BC^\times$.

\subsubsection{Elliptic vertex operators}\label{sec:elliptic_VO}

We define the elliptic deformation of the vertex operators using the doubling trick~\cite{Clavelli:1973uk},
\begin{subequations}
\begin{align}
 \sA_i(x) & = \ :\exp \left( a_{i,0} + \sum_{n \neq 0} a_{i,n}^{(+)} x^{-n} + \sum_{n \neq 0} a_{i,n}^{(-)} x^{+n} \right):
 \, , \\
 \sY_i(x) & = \ :\exp \left( y_{i,0} + \sum_{n \neq 0} y_{i,n}^{(+)} x^{-n} + \sum_{n \neq 0} y_{i,n}^{(-)} x^{+n} \right):
 \, ,
\end{align}
\end{subequations}
where the oscillators obey the commutation relations,
\begin{subequations}
\begin{align}
 \left[ a_{i,n}^{(\pm)} , \, a_{j,m}^{(\pm)} \right] & = \mp \frac{1}{n} \frac{(1 - q_1^{\pm n})(1 - q_2^{\pm n})}{1 - p^{\pm n}} \, c_{ji}^{[\pm n]} \, \delta_{n+m,0}
 \, , \\
 \left[ y_{i,n}^{(\pm)} , \, y_{j,m}^{(\pm)} \right] & = \mp \frac{1}{n} \frac{(1 - q_1^{\pm n})(1 - q_2^{\pm n})}{1 - p^{\pm n}} \, \tilde{c}_{ij}^{[\mp n]} \, \delta_{n+m,0}
 \, .
\end{align}
\end{subequations}
These vertex operators provide elliptic analog of the OPE factors in Sec.~\ref{sec:OPEs}, replacing all the factors with the elliptic function.
The $\sV$-operator is similarly defined to provide the elliptic OPE factor,
\begin{subequations}
 \begin{align}
  \sV_i(x) \sA_j(x') & = \theta\left( \frac{x'}{x};p \right)^{-\delta_{ij}} :\sV_i(x) \sA_j(x'): \, ,
  \\
  \sA_j(x') \sV_i(x) & = \theta\left( \frac{x}{x'};p \right)^{-\delta_{ij}} :\sV_i(x) \sA_j(x'): \, .
 \end{align}
\end{subequations}

In this case, the Fock space is generated by two sets of the formal parameters, $(t_{i,n}^{(\pm)}, \partial_{i,n}^{(\pm)})_{i \in \Gamma_0, n = 1,\ldots, \infty}$, which is a direct product of the Fock spaces generated by the plus and minus sectors.
For example, the vacuum is given by $\ket{0} = \ket{0}^{(+)} \oplus \ket{0}^{(-)}$.
Then we obtain exactly the same formula of the $k$-instanton contribution as \eqref{eq:Zk_correlator}, and the total partition function \eqref{eq:Z_correlator}, as a correlator with respect to the doubled Fock space equipped with the elliptic vertex operators defined here.

\subsection{$qq$-character}

We can similarly consider the $qq$-character in 6d $\CalN = (1,0)$ theory on $\BC^2 \times T^2$.
Since the $qq$-character discussed in this paper is an operator, it is not expressed as a correlator, and we should use the elliptic vertex operator formalism used in Sec.~\ref{sec:elliptic_VO} to construct the $qq$-character in this case.
Using the elliptic vertex operators with the doubled Fock space, we obtain formally the same formula as \eqref{eq:T_operator}, and also \eqref{eq:T_operator_full}, for arbitrary highest-weight modules constructed on a quiver.

%%% References

\bibliographystyle{amsalpha_mod}
%\bibliographystyle{utphys}
%\bibliography{../wquiver/wquiver}
\bibliography{/Users/k_tar/Documents/Repository/wquiver/wquiver}
%\bibliography{/Users/k_tar/Dropbox/etc/conf}

\end{document}